\documentclass{ws-procs9x6}

\newcommand{\beq}{\begin{equation}}
\newcommand{\eeq}{\end{equation}}
\newcommand{\beqs}{\begin{eqnarray}}
\newcommand{\eeqs}{\end{eqnarray}}

\begin{document}

\title{Study of the Change from Walking to Non-Walking Behavior\\
in a Vectorial Gauge Theory as a Function of $N_f$
}

\author{M. KURACHI$^*$ and R. SHROCK$^\dagger$}

\address{C.N. Yang Institute for Theoretical Physics,\\
State University of New York,\\
Stony Brook, New York 11794, USA\\
$^* speaker; $E-mail: masafumi.kurachi@sunysb.edu\\
$^\dagger$E-mail: robert.shrock@sunysb.edu}

\begin{abstract}
Based on recent works \cite{Kurachi:2006ej, Kurachi:2006mu}, we present the
results of calculations for several physical quantities (meson masses, the $S$
parameter, etc.)  in a vectorial gauge theory, as a function of the number of
fermions, $N_f$. Solutions of the Schwinger-Dyson and the Bethe-Salpeter
equations with the improved ladder approximation are used for the
calculations. We focus on how the values of physical quantities change as one
moves from the QCD-like (non-walking) to walking regimes.
\end{abstract}

\bodymatter

\section{Introduction}

We consider a $(3+1)$-dimensional vectorial gauge theory (at zero temperature
and chemical potential) with the gauge group SU($N_c$) and $N_f$ massless
fermions transforming according to the fundamental representation of this
group.  For $N_c=3$, if one took $N_f=2$, this would be an approximation to
actual QCD.  We restrict here to the range $N_f < (11/2)N_c$ for which the
theory is asymptotically free.  An analysis using the two-loop beta function
and Schwinger-Dyson equation leads to the inference that for $N_f$ in this
range, the theory includes two phases: (i) for $0 \le N_f \le N_{f,cr}$ a phase
with confinement and spontaneous chiral symmetry breaking (S$\chi$SB); and (ii)
for $N_{f,cr} \le N_f \le (11/2)N_c$ a phase with no spontaneous chiral
symmetry breaking (plausibly a non-Abelian Coulomb phase). We shall refer to
$N_{f,cr}$, the critical value of $N_f$, as the boundary between these two
phases.

For $N_f$ slightly less than $N_{f,cr}$, the theory exhibits an approximate
infrared (IR) fixed point with resultant walking behavior.  That is, as the
energy scale $\mu$ decreases from large values, $\alpha=g^2/(4\pi)$ ($g$ being
the SU($N_c$) gauge coupling) grows to be O(1) at a scale $\Lambda$, but
increases only rather slowly as $\mu$ decreases below this scale, so that there
is an extended interval in energy below $\Lambda$ where $\alpha$ is large, but
slowly varying.  Associated with this slowly running behavior, the resultant
dynamically generated fermion mass, $\Sigma$, is much smaller than $\Lambda$.
In addition to its intrinsic field-theoretic interest, this walking behavior
has played an important role in theories of dynamical electroweak symmetry
breaking \cite{wtc1}-\cite{chipt3}.  As $N_f$ approaches $N_{f,cr}$ from below,
quantities with dimensions of mass vanish continuously; i.e., the chiral phase
transition separating phases (i) and (ii) is continuous.  Recently, meson
masses and other quantities such as the generalized pseudoscalar decay constant
$f_P$ and the $S$ parameter \cite{pt} were calculated in the walking limit of
an SU($N_c$) gauge theory\cite{mm,HKY}.

It is of interest to investigate how meson masses and other quantities change
as one decreases $N_f$ below $N_{f,cr}$, moving away from the boundary, as a
function of $N_f$, between phases (i) and (ii), deeper into the confined phase.
For this purpose, in Ref. \cite{Kurachi:2006ej}, as in Refs. \cite{mm, HKY}, we
use the Schwinger-Dyson (SD) equation to compute the dynamical fermion mass
$\Sigma$ (generalized constituent quark mass) and then insert this into the
Bethe-Salpeter (BS) equation to obtain the masses of the low-lying mesons and
other quantities.  We restrict to an interval of $N_f$ values for which the
theory has an infrared fixed point. For definiteness, we take $N_c=3$; however,
$N_c$ enters only indirectly, via the dependence of the value of the infrared
fixed point $\alpha_*$ on $N_c$.  Hence, our findings may also be applied in a
straightforward way, with appropriate changes in the value of $\alpha_*$, to an
SU($N_c$) gauge theory with a different value of $N_c$.

In order to study meson masses and other quantities as one moves away from the
boundary between phases (i) and (ii), it is first necessary to know as
accurately as possible where this boundary lies, as a function of $N_f$, i.e.,
to know the value of $N_{f,cr}$.  For sufficiently large $N_f$, the beta
function (calculated to the maximal scheme-independent order, namely two loops)
has an IR fixed point at
\beq
\alpha_* = \frac{-4\pi(11N_c -2N_f)}{34N_c^2-13N_cN_f+3N_c^{-1}N_f} \ . 
\label{alpha_irfp}
\eeq
Requiring that $\alpha_*$ be sufficiently large as to yield spontaneous
symmetry breaking in the context of an approximate solution to the 
SD equation for a fermion 
yields the condition that $N_f < N_{f,cr}$, where \cite{chipt3} 
\beq
N_{f,cr} = \frac{2N_c(50N_c^2-33)}{5(5N_c^2-3)} \ . 
\label{nfcr}
\eeq
For $N_c=3$ this gives $N_{f,cr} \simeq 11.9$.  These estimates are only rough,
in view of the strongly coupled nature of the physics.  Effects of higher-order
gluon exchanges and instantons have been studied in Refs. \cite{alm}.  

In our analysis, what we actually vary is the value of the approximate IR fixed
point $\alpha_*$, which depends parametrically on $N_f$.  Thus, although our SD
and BS equations are semi-perturbative, the analysis is self-consistent in the
sense that our $\alpha_{cr}$ really is the value at which, in our
approximation, one passes from the confinement phase with S$\chi$SB to the
chirally symmetric phase, and our values of $\alpha_*$ do span the interval
over which there is a crossover from walking to QCD-like (i.e., non-walking)
behavior.

\section{Schwinger-Dyson Equation}

\begin{figure}[t]
  \begin{center}
    \includegraphics[height=3.5cm]{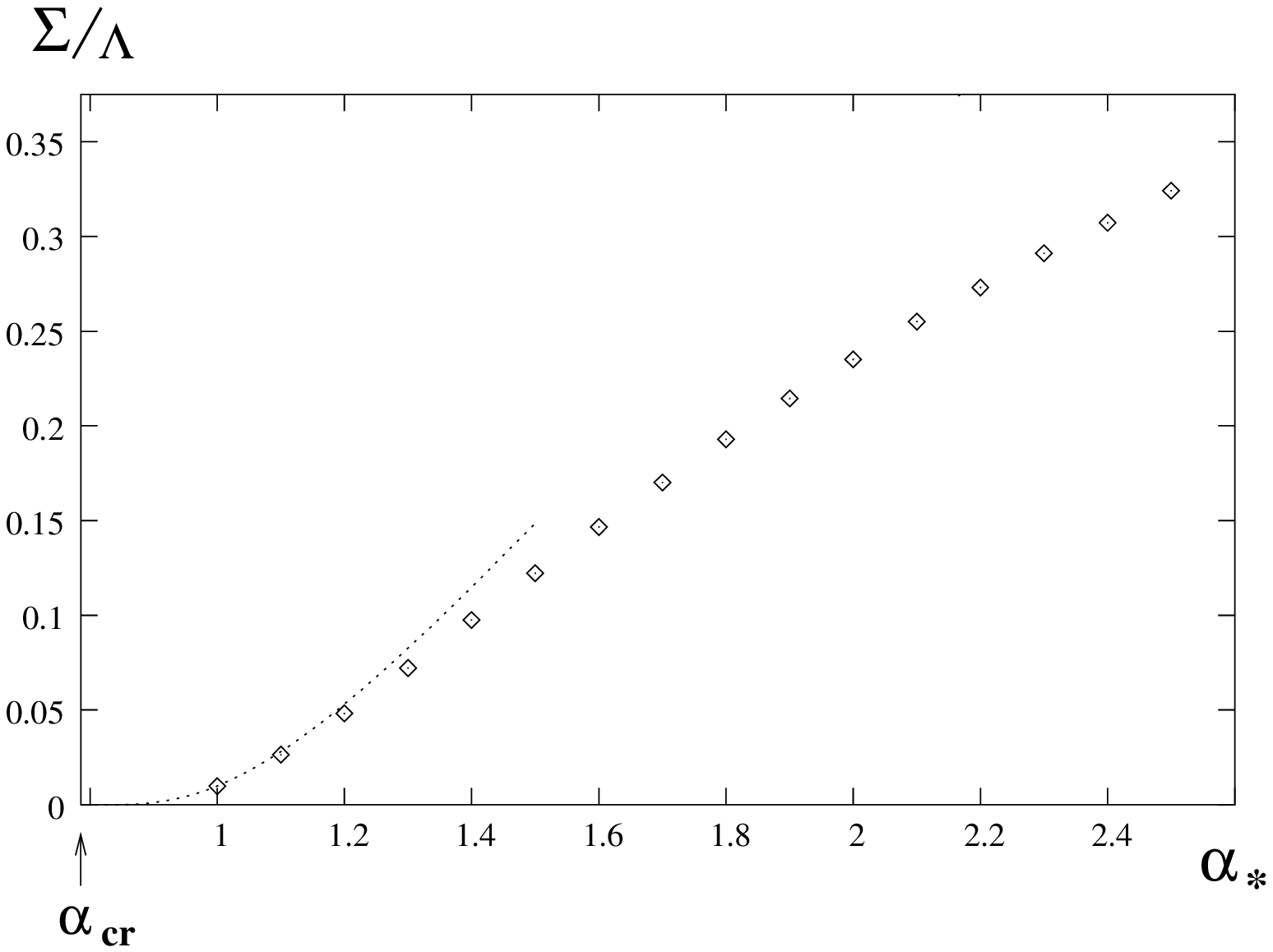}
    \includegraphics[height=3.5cm]{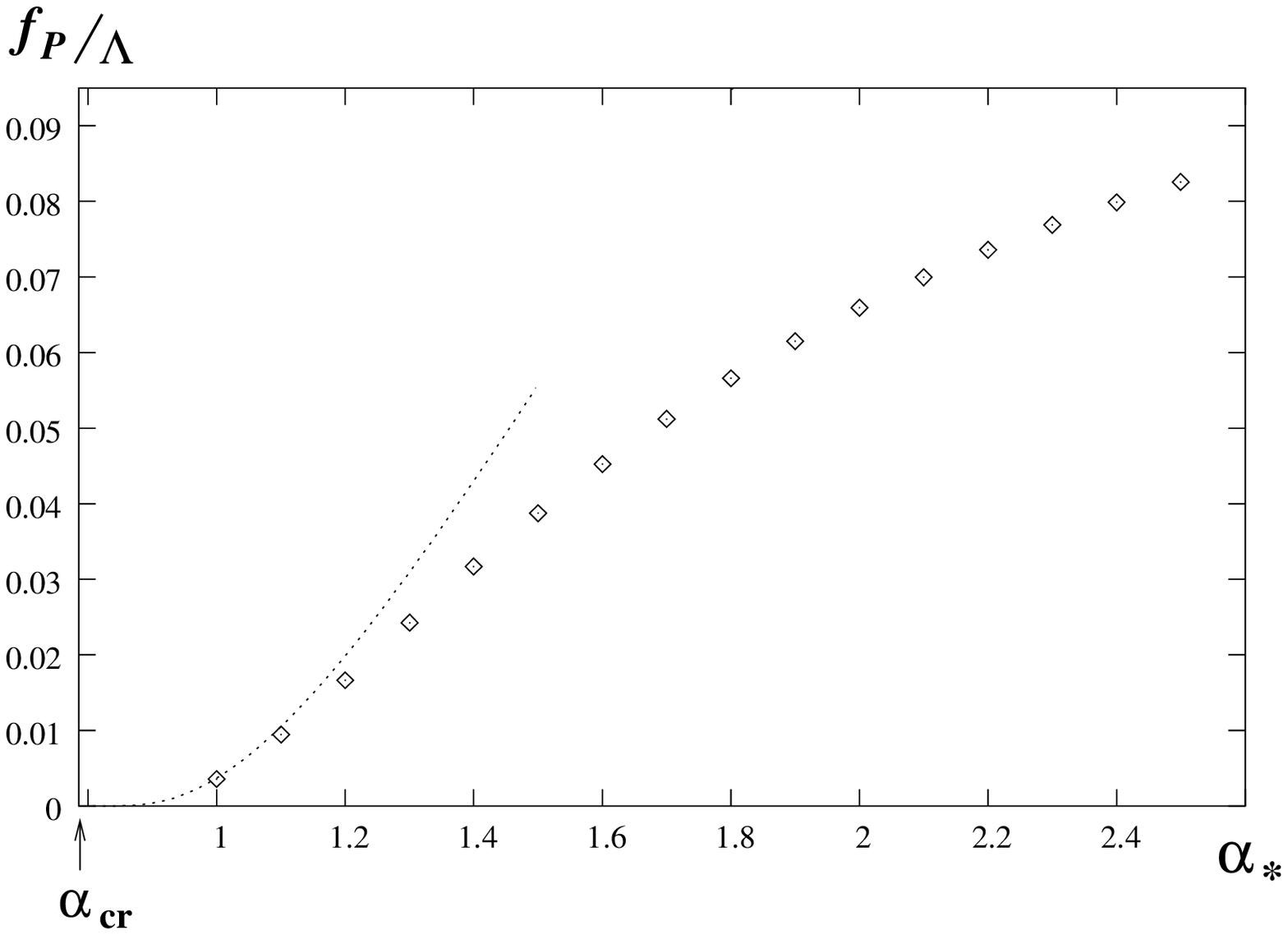}
  \end{center}
\caption{Numerical solutions for $\Sigma$ (left panel) and 
$f_P$ (right panel), for several values of $\alpha_*$
(indicated by $\Diamond$). For comparison, we show eq. (\ref{sigsol}) with 
$c=4.0$ (left panel), $c=1.5$ (right panel) from a fit 
to the results for $0.89 \le \alpha_* \le 1.0$.}
\label{Sigma_comp}
\end{figure}

We first use the Schwinger-Dyson equation for the fermion propagator to
calculate the dynamically generated mass $\Sigma$ of this fermion.  In
Fig. \ref{Sigma_comp} (left panel) 
we show the solution for the dynamical fermion mass
$\Sigma$ as a function of $\alpha_*$.  A fit to the numerical solution in the
walking region $0.89 \le \alpha_* \le 1.0$ \cite{mm} found agreement
with the functional form
\beq
\Sigma = c \Lambda \, \exp
\bigg [ -\pi \Big ( \frac{\alpha_*}{\alpha_{cr}} - 1 \Big )^{-1/2} \bigg ] \ ,
\label{sigsol}
\eeq
with $c = 4.0$ (see also Refs. \cite{chipt2,chipt3}). Our calculations for
larger $\alpha_*$ show the expected shift away from walking behavior.  This
shift is evident in Fig. \ref{Sigma_comp} for $\alpha_*$ larger than about 1.2.
Our calculation of $\Sigma$, shown in Fig. \ref{Sigma_comp}, shows that
$\Sigma/\Lambda$ increases substantially, by about a factor of 30, from a value
of about 0.01 at $\alpha_* = 1.0$ to 0.32 at $\alpha_*=2.5$, much closer to the
value of O(1) for this ratio in QCD.

Another quantity of interest is the pseudoscalar decay constant $f_P$, the
$N_f$-flavor generalization of the pion decay constant, $f_\pi$.  In
Fig. \ref{Sigma_comp} (right panel) we show our results for $f_P$ calculated
by substituting our solution for $\Sigma(k^2)$ into the Pagels-Stokar
formula. In the walking limit, $f_P$ has been shown to satisfy a relation
similar to eq. (\ref{sigsol}), i.e., it is exponentially smaller than the scale
$\Lambda$.  We display, as the dotted curve, the fit from Ref. \cite{mm} for
the walking interval $0.89 \le \alpha_* \le 1.0$, given by eq. (\ref{sigsol})
with $c=1.5$.  Our results show the change from this walking type of behavior
as $\alpha_*$ increases above this range; specifically, as $\alpha_*$ increases
from 1.0 to 2.5, $f_P/\Lambda$ increases substantially, from about $ 3 \times
10^{-3}$ to about 0.08.  This is similar to the factor by which we found that
$\Sigma/\Lambda$ increased as $\alpha_*$ increased through this interval.

\section{Calculation of Meson Masses}

\begin{figure}[t]
  \begin{center}
    \includegraphics[height=4.7cm]{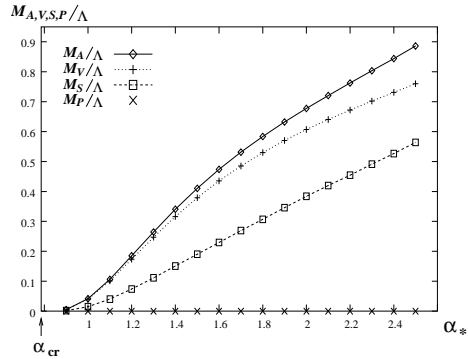}
  \end{center}
\caption{Values of meson masses divided by $\Lambda$ calculated from the
 Schwinger-Dyson and Bethe-Salpeter equations 
 in the range $0.9 \le \alpha_* \le 2.5$.}
\label{m_over_lam}
\end{figure}

\begin{figure}[t]
  \begin{center}
    \includegraphics[height=4.8cm]{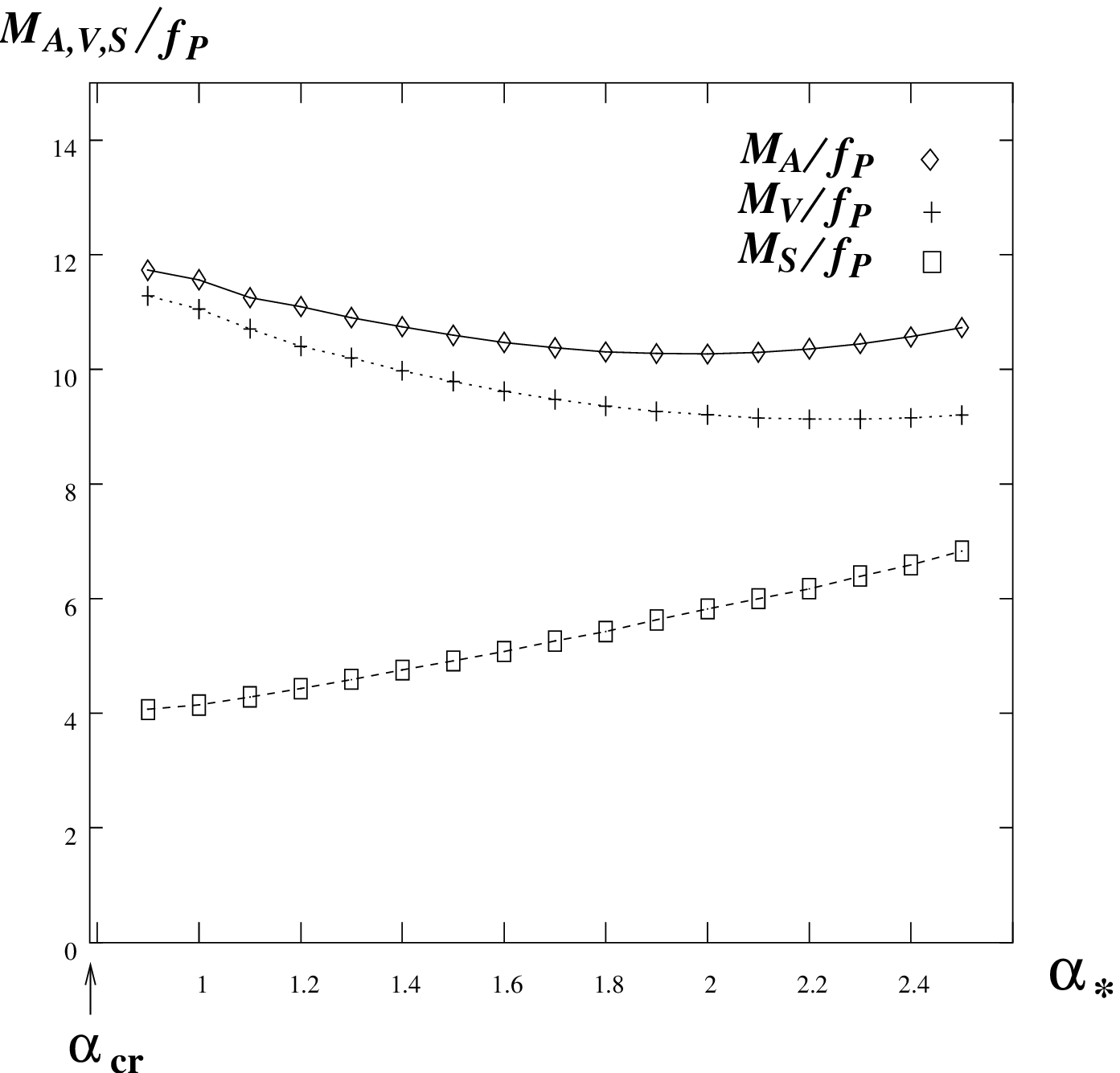}
    \includegraphics[height=4.8cm]{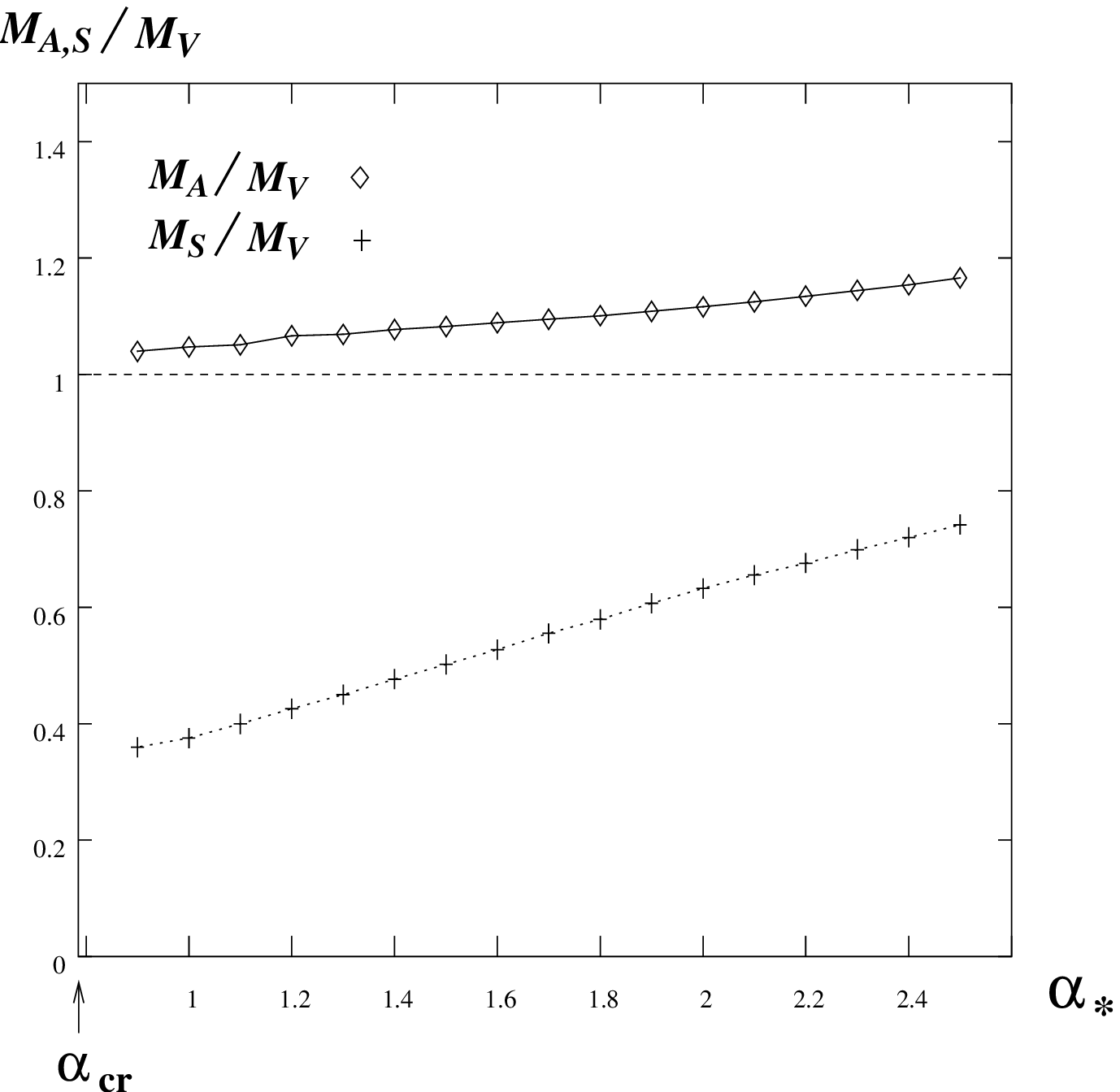}
  \end{center}
\caption{Values of meson masses divided by $f_P$ (left panel) 
and $M_{A, S}/M_V$ (right panel) calculated from the
 Schwinger-Dyson and Bethe-Salpeter equations.}
\label{m_over_fp}
\end{figure}

We next present the results of the numerical calculations for meson masses,
obtained by solving the homogeneous BS equation
\cite{Kurachi:2006ej}.  (As in Ref. \cite{mm}, we have checked and
confirmed that the flavor-adjoint pseudoscalar meson mass is zero to within the
numerical accuracy of our calculation.) In Fig.~\ref{m_over_lam}, we show the
values of meson masses divided by $\Lambda$ calculated from the SD and BS
equations.  In Fig.~\ref{m_over_fp} we plot the values of $M_{A,V,S}/f_P$ (left
panel) and $M_{A, S}/M_V$ (right panel).  Here, the subscripts $V, A$ and $S$
represent vector, axial-vector and scalar, respectively.


Our calculations yield a number of interesting results.  We summarize these for
the changes in these meson masses as $\alpha_*$ increases from 0.9 to 2.5.  The
ratios of the meson masses divided by $\Lambda$ increase dramatically, by
factors of order $10^2$, approaching values of order unity at $\alpha_* = 2.5$.
This amounts to the removal of the exponential suppression of these masses
which had described the walking limit near $N_{f,cr}$, as one moves away from
this limit into the interior of the confined phase.  For example, $M_S/f_P$
increases monotonically from about 4 to 7, thereby approaching to within about
35 \% of the value 10.7 in QCD for $M_{a_0}/f_\pi$, while $M_V/f_P$ decreases
from about 11 to 9, rather close to the value 8.5 for $M_\rho/f_\pi$ and
$M_\omega/f_\pi$ in QCD.  The ratios $M_A/M_V$ and $M_S/M_V$, which were found
in Ref. \cite{mm} to have values close to 1.0 and 0.36, respectively, in the
walking limit, both increase in the interval of $\alpha_*$ that we study,
reaching about 1.2 and 0.74, respectively, at $\alpha_*=2.5$.  For comparison,
these ratios are approximately 1.6 and 1.3 in QCD.

\section{Calculation of the $S$ parameter}

In this section we present the results of our calculations of $\hat{S}$
\cite{Kurachi:2006mu} in the crossover region of the theory between the walking
limit at $\alpha_* \searrow \alpha_{cr}$ and larger values of $\alpha_*$ that
move toward the QCD-like regime. Here, $\hat{S}$ represents the contribution to
the $S$ parameter from one fermion isodoublet.  (Studies of $S$ in the walking
limit include \cite{HKY,scalc}.)  We calculate $\hat{S}$ via the relation
$\hat{S}=4\pi (\Pi_{VV}'(0)-\Pi_{AA}'(0))$, where $\Pi_{VV}(q^2)$ and
$\Pi_{AA}(q^2)$ are the vector and the axial-vector current-current correlation
functions. These correlators are computed by solving the SD equation and the
inhomogeneous BS equation \cite{Kurachi:2006mu}.

\begin{figure}[t]
  \begin{center}
    \includegraphics[height=5cm]{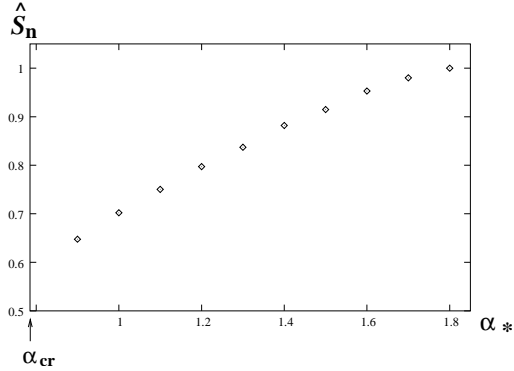}
  \end{center}
\caption{Plot of $\hat{S}_n$ for several values of $\alpha_\ast$ in 
the range of $0.9\le \alpha_\ast \le 1.8$. As indicated by the subscript 
$n$, the values are normalized by the value of $\hat{S}$ at 
$\alpha_\ast=1.8$, i.e., 0.47.}
\label{S-hat_n}
\end{figure}

In Fig.~\ref{S-hat_n}, as a function of $\alpha_\ast$, we plot the value of 
$\hat{S}_n$, the value of $\hat S$ normalized by its value at 
$\alpha_\ast=1.8$, namely, 0.47.  
This figure shows that $\hat{S}_n$, and hence also $\hat S$, decreases by about
40 \% as $\alpha_*$ is reduced from 1.8 to 0.9, or equivalently as $N_f$ is
increased from 10.3 to 11.6.  Reinserting the factor of the number of fermion
isodoublets, $N_D=N_f/2$, to get $S$ itself, we obtain a decrease by about 30
\% in $S$, since $N_D$ only increases by about 10 \% over this range.  Thus,
our calculation shows that for this range of values, $S$ decreases
significantly as one moves from the QCD-like to the walking regimes.  We recall
that the (improved) ladder approximation to the BS equation can 
overestimate $S$ in QCD by as much as 30 \% \cite{hys}.  Hence, in addition to
the demonstrated decreasing trend of $\hat S$ and $S$ as $\alpha_*$ decreases
from 1.8 to 0.9, one may, separately, comment that the absolute magnitude of
these quantities could be about 30 \% smaller than the values yielded by our
ladder approximation.  Our results thus strengthen the evidence for the
reduction of $\hat S$ in a walking, as opposed to QCD-like, gauge theory, and
are relevant to assessing the impact of the $S$-parameter constraint on
technicolor theories.

\section{Summary}
In summary, using numerical solutions of the Schwinger-Dyson and Bethe-Salpeter
equations, we have calculated several physical quantities, including $f_P$,
meson masses, and the $S$ parameter, as a function of the approximate
infrared fixed point, $\alpha_*$, or equivalently, the number of massless
fermions, $N_f$, in a vectorial, confining SU($N$) gauge theory.  Our results
show the crossover between walking and non-walking behavior in a gauge theory,
and demonstrate that $\hat S$ and also $S$ decrease significantly as $\alpha_*$
decreases in this range.

\section*{Acknowledgments}
M.K. thanks Profs. M. Harada and K. Yamawaki for the collaborations on the
related Refs. \cite{mm,HKY}.  
This research was partially supported by the grant NSF-PHY-03-54776.


\begin{thebibliography}{40}

%\cite{Kurachi:2006ej}
\bibitem{Kurachi:2006ej}
  M.~Kurachi and R.~Shrock,
  %``Study of the change from walking to non-walking behavior in a vectorial
  %gauge theory as a function of N(f),''
  JHEP {\bf 0612}, 034 (2006)
  %%CITATION = HEP-PH 0605290;%%

%\cite{Kurachi:2006mu}
\bibitem{Kurachi:2006mu}
  M.~Kurachi and R.~Shrock,
  %``Behavior of the S parameter in the crossover region between walking and
  %QCD-like regimes of an SU(N) gauge theory,''
  Phys.\ Rev.\ D {\bf 74}, 056003 (2006)
  %%CITATION = HEP-PH 0607231;%%

\bibitem{wtc1}
B. Holdom, Phys. Lett. B {\bf 150}, 301 (1985).

\bibitem{wtc2} 
K. Yamawaki, M. Bando, and K. Matumoto, Phys. Rev. Lett. {\bf
56}, 1335 (1986).

\bibitem{chipt1}
T. Appelquist, D. Karabali, and L. C. R. Wijewardhana, Phys. Rev. Lett. {\bf
57}, 957 (1986); T. Appelquist and L. C. R. Wijewardhana, Phys. Rev. D
{\bf 35}, 774 (1987); Phys. Rev. D {\bf 36}, 568 (1987).

\bibitem{chipt2}
T. Appelquist, J. Terning, and L. C. R. Wijewardhana,
Phys. Rev. Lett.  {\bf 77}, 1214 (1996).

\bibitem{my}
V. Miransky and K. Yamawaki, Phys. Rev. D {\bf 55}, 5051 (1997); {\it ibid.} 
{\bf 56}, E 3768 (1997). 

%\cite{Chivukula:1996kg}
\bibitem{Chivukula:1996kg}
  R.~S.~Chivukula,
  %``A comment on the zero temperature chiral phase transition in SU(N)  gauge
  %theories,''
  Phys.\ Rev.\ D {\bf 55}, 5238 (1997)
  %%CITATION = HEP-PH 9612267;%%

\bibitem{chipt3} 
T. Appelquist, A. Ratnaweera, J. Terning, and 
L. C. R. Wijewardhana, Phys. Rev. D {\bf 58}, 105017 (1998).

\bibitem{pt}
M.~E.~Peskin and T.~Takeuchi, Phys.\ Rev.\ Lett.\  {\bf 65}, 964 (1990);
Phys.\ Rev.\ D {\bf 46}, 381 (1992).

\bibitem{mm}
% meson masses in large N_f QCD 
M. Harada, M. Kurachi, and K. Yamawaki, Phys. Rev. D {\bf 68}, 076001 (2003).

\bibitem{HKY}
M. Harada, M. Kurachi, and K. Yamawaki, Prog. Theor. Phys. {\bf 115}, 765 
(2006). 

\bibitem{alm} 
T. Appelquist, K. Lane, and U. Mahanta, Phys. Rev. Lett. {\bf
61}, 1553 (1988); T. Appelquist et al., Phys. Rev. D {\bf 43}, 646 (1991); 
T. Appelquist and S. Selipsky, Phys. Lett. B {\bf 400}, 364 (1997).

\bibitem{scalc}
T. Appelquist and G. Triantaphyllou, Phys. Lett. B {\bf 278}, 345 (1992);
R. Sundrum and S. Hsu, Nucl. Phys. B {\bf 391}, 127 (1993);
T. Appelquist and F. Sannino, Phys. Rev. D {\bf 59}, 067702 (1999);
S. Ignjatovic, L. C. R. Wijewardhana, and T. Takeuchi, Phys. Rev.
D {\bf 61}, 056006 (2000).

\bibitem{hys}
M.~Harada, M. Kurachi, and K. Yamawaki, Phys. Rev. D {\bf 70}, 033009 (2000).

\end{thebibliography}
\end{document}